# Radio frequency signal detection by ballistic transport in Y-shaped graphene nanoribbons

G.Deligeorgis [1,2] *, F.Coccetti [1,2], G. Konstantinidis [3] and R.Plana [1,2]

[1] *CNRS, LAAS, 7 avenue du colonel Roche, F-31400 Toulouse, France*

[2] *Univ de Toulouse, LAAS, F-31400 Toulouse, France.*

[3] *Foundation for Research & Technology Hellas (FORTH) P.O. BOX 1527, Vassilika Vouton, Heraklion 711 10, Crete, Greece*

We report on the fabrication and room temperature measurements of a high frequency electrical signal detector. The device is based on the ballistic transport in graphene to detect a high frequency signal. The observed response is linear in the considered power range (-40dBm – 0dBm) and exhibits a sensitivity as high as 10 Volts/Watt. Finally the device detected signals up to 50GHz with a maximum response at 10GHz. This device outperforms any previously reported carbon based detectors in frequency response and compares to current state of the art Schottky based detectors in dynamic range.

Graphene is attracting considerable attention in the past ten years, since its isolation by A. Geim and K. Novoselov [1]. The rates at which achievements are made public is astonishing. Although in electronics the main focus was towards transistors [2–4] where ultra fast operating speeds were promised, graphene has also surfaced at the epi-center of recent advances in sensing. Specifically chemical sensing [5], DNA sequencing [6], [7] and light detectors [8] operating in the GHz range have been demonstrated by exploiting the unique properties of the material. Graphenes' ability to transfer carriers with little or no scattering over a large distance [9], [10], has been proposed as a major feature for spintronics [11], [12] as well as for even more exotic effects such as the "Veselago" lensing effect [13]. Indeed graphene routinely exhibits a mean free path in the order of 300nm [14] which is significantly larger than the minimum feature technologically feasible in modern electronics. Until recently however, this remarkable feature, was not utilized in applications. This property can be used to obtain a group of devices where the transport phenomena are governed not by drift and diffusion but by ballistic transport of carriers [15], [16].

Radio frequency (RF) detectors have numerous applications ranging from metrology to RF demodulation for telecommunications. Numerous RF detection technologies exist and are based on several different techniques. The most widely used today are micro-bollometers [17] and Schottky diodes [18]. Each technique has different advantages and disadvantages for detection in terms of linearity, frequency cut-off, minimum detected power and the ability to detect fast signals (important for reception in telecommunications). As far as carbon related RF detectors, few reports exist in the literature and they predominantly focus on rectifying contacts fabricated by carbon nanotubes [19].

---
\* Corresponding author: gdeligeo@laas.fr

In this work, we present our attempt to take advantage of the ballistic transport in graphene to create a device able to convert a high frequency signal into a low frequency electrical output.

The fabricated device consists of two high frequency ground – signal – ground ports between which, a coplanar waveguide structure is formed. The ground electrodes are continuous between the two ports. The right side (RF input port) leads to an open stub geometry. The gap between the signal and ground electrodes is ~2μm (Fig.1). Inside the signal to ground electrode gap, a series of "Y" shaped graphene nanoribbons are formed by Oxygen plasma etching. The left top and right top edges of each nanoribbon are electrically in contact with the signal and the ground electrode respectively. To contact the bottom edge of each nanoribbon, without contacting the center of the "Y" shaped graphene junction, a metallic air-bridge is formed between adjacent bottom electrodes. This metallic air-bridge is anchored at the bottom of each "Y" and eventually is electrically connected to the signal electrode that leads to the output port (left port in Fig.1). The patterns were fabricated by electron beam lithography on PMMA resist and all electrodes were realized by evaporation of a 2nm Cr / 250nm Au using an electron gun evaporator. Following lift-off of the excess metal, no other treatment was carried out on the device (no thermal or current annealing and no post-processing cleaning treatment).

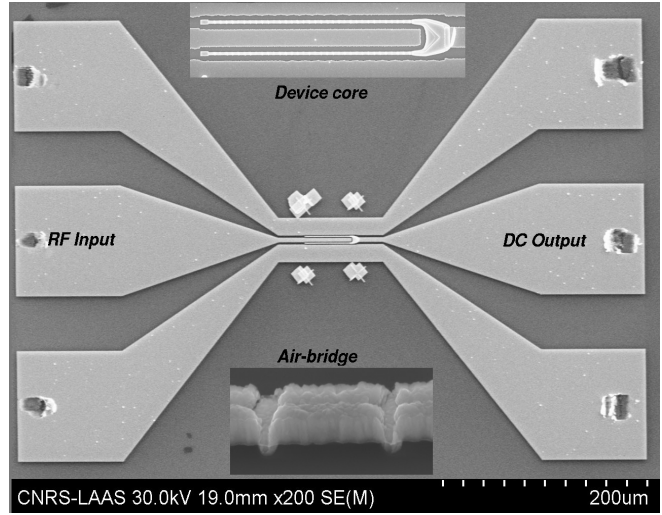

*Figure 1: Top-view image of the fabricated device. The top inset shows the core of the device with more detail. The bottom inset show a detail of the air-bridge used to form the central electrode. The Y-shaped graphene nanoribbons reside below the suspended part of the air-bridge and connect to the center electrodes, the ground and the feet of the air-bridge structure. Thus a parallel action from all nanoribbons is obtained.*

When an RF signal is applied to the input port, an alternating field formed in the gap of the coplanar waveguide will be applied to the nanoribbon junctions. The electric field will excert a force on carriers inside the graphene nanoribbon top arms. Should each nanoribbon junction be modeled by a star like resistance network, the response in the bottom electrode would be the sinusoidal input signal divided in amplitude by the access resistances corresponding to the two top arms that are connected to the signal and the ground electrodes respectively. Taking however into account the fact that ballistic transport may occur, a non linear response is expected [20].

To measure the device response, a high frequency signal generator was connected to the input port (Agilent 8257D). Input power levels between -40 and 0 dBm were used. Input frequency was varied from 50MHz up to 50GHz. On the output port, a spectrum analyzer (ROHDE&SWARTZ FSU) and a Digital Multimeter (Keithley DMM4200) unit were connected using a T-bias to separate low and high frequency signals. All measurements were performed at room temperature. The described setup enabled us to apply any given RF-signal to the device under test and measure the transmission in high frequency as well as the DC response of the device. High frequency losses were measured for the input port network in order to estimate the true power delivered to the device. All power values presented in this letter are corrected for this effect unless otherwise noted.

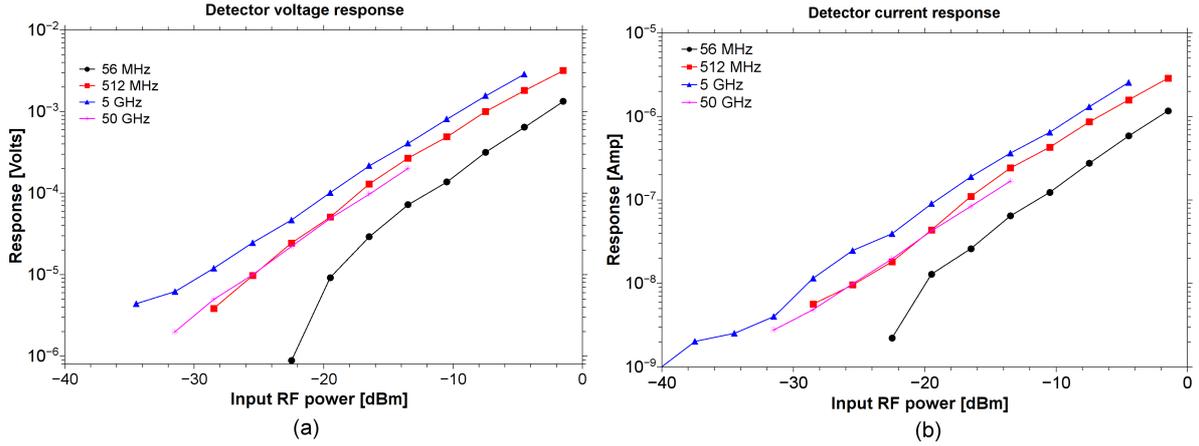

*Figure 2: Open circuit voltage (a) and current (b) response of the detector as a function of input power. The four curves correspond to different input frequencies.*

The first experiment was to measure the response of the device as a function of input power. For each power delivered to the device the DC response was measured by the Keithley DMM connected to the output. Both the "Open circuit voltage" $V_{oc}$ and the "Short circuit current" $I_{sc}$ at the output port were measured. The investigated power range was varied from 0dBm down to -40dBm (RF source power). This experiment was repeated for four different input frequencies ranging from 50MHz to 50GHz.

The measured voltage response of the device is shown in a log-log plot for four different frequencies (Fig.2a). From fig.2 we can verify from the slope in the log-log plot that the response depends linearly with power for the measured power range which corresponds to four orders of magnitude. This is verified for all measured frequencies. The $I_{sc}$ follows a similar response in the range of 1nA to several uA. The corresponding plot is shown in Fig.2b. Plotting the data shown in Fig.2 in a linear plot (response as a function of power in Watts), the sensitivity of the device can be calculated. A response as high as 10Volts/Watt (at 10 GHz) was obtained.

We argue that such a response may not be explained by assuming a resistive behavior of the graphene nanoribbon junctions. Instead, a possible explanation is the ballistic transport of carriers in the junction. Carriers forced to move by the electric field present in the gap of the waveguide do so in the upper branches of the junction, some of them are forced towards the junction where they will either traverse the junction area without scattering – and enter the bottom arm – or they will scatter and move governed by drift. The carriers that do not scatter, will end up at the bottom arm where they may be collected by the output electrode. This process does not depend on the sign of the electric field in the gap thus a rectification effect is developed. Such argument is supported by the fact that the total size of the junction is similar to the mean free path typically measured in graphene monolayers.

To accurately evaluate the frequency response of the device, the sensitivity was extracted as described in the previous paragraph for several frequencies. The sensitivity vs frequency is plotted in Fig.3. The response of the detector as a function of frequency can be divided in three regions. In the first, which corresponds to a frequency from 50MHz to 1GHz, the response is increasing. In the second part of the measured frequencies namely from 1GHz to 10GHz the response is marginally increasing with the measured frequency. Finally in the range of 10GHz and upwards, the response of the device is declining. The same behavior is exhibited for both voltage and current response (not shown here).

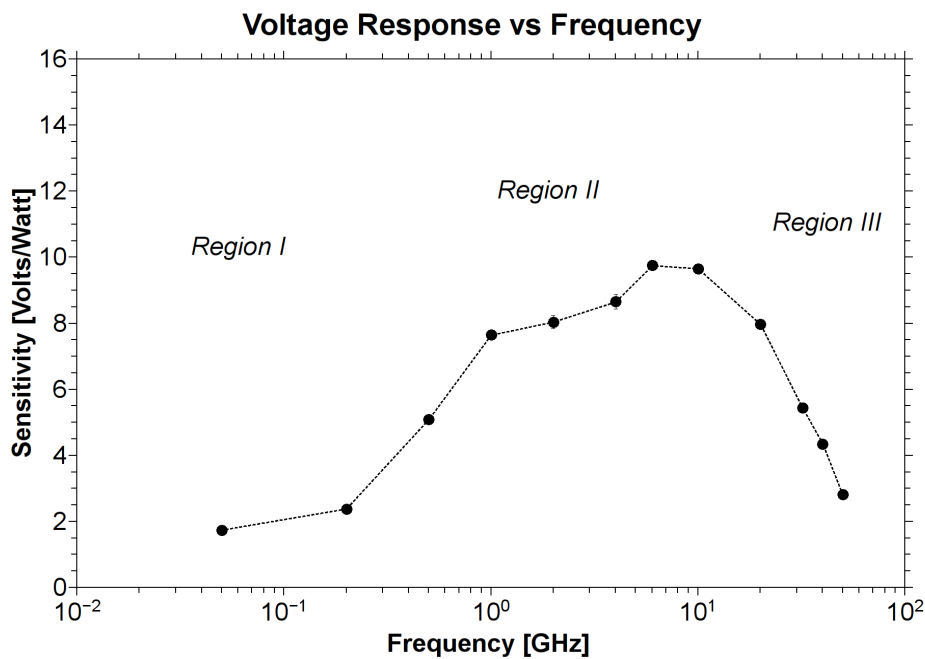

*Figure 3: Sensitivity as a function of input frequency. A maximum that corresponds to 10Volts/Watt was measured for 10 GHz and a cut-off frequency of 35 GHz. Low frequency losses can be attributed to impedance mismatch evident in scattering parameter measurements (not shown here)*

To explain the frequency dependent response of the device, we should take a closer look at the RF behavior of the input port. Indeed for low frequency, a mismatch in impedance causes a significant input reflection loss which translates to reduced power at the core of the device where the nanoribbons reside. This phenomenon is particularly pronounced up to 1

GHz in the scattering matrix coefficients as measured using a vector network analyzer (VNA). This frequency coincides with the limit between regions I and II as identified in Fig3. At the high frequency limit, a significant reduction in response is exhibited. The origins of this cut-off frequency are still under investigation. Assuming a -3dB cut-off level compared to the maximum performance a cut-off frequency of 35GHz is calculated.

In conclusion we have designed, fabricated and characterized a RF signal detector based on graphene. The device is operating efficiently as a detector exhibiting a linear response as high as 10Volts/Watt over a large dynamic range and is able to detect signals at least as low as -40dBm (100nWatt). The measured performance is superior to any carbon based passive RF detector presented so far. Indications that the detector has a particularly fast response time pave the way towards RF applications of graphene based detectors of this kind. Such a device may significantly simplify the existing receiver circuitry used in modern electronics.

**Acknowledgments**

All authors would like to acknowledge support by the LEA-smartmems joint European laboratory. G. Konstantinidis would also like to acknowledge support by the Greek Graphene center.